\documentclass[10pt,conference,letterpaper]{IEEEtran}

\usepackage{cite}
\usepackage{amsmath,amssymb,amsfonts}
\usepackage{graphicx}
\usepackage{textcomp}
\usepackage{xcolor}
\usepackage{microtype}
\usepackage{tikz}
\usetikzlibrary{positioning,arrows.meta}

\usepackage{url}
\usepackage[hidelinks]{hyperref} 
\usepackage{tabularx}
\usepackage{etoolbox}

\usepackage{enumerate}
\usepackage{comment}

\usepackage{mathtools}

\usepackage{texdef2015}

\newcommand{\tabref}[1]{Table~\ref{tab:#1}}
\newcommand{\secref}[1]{Section~\ref{sec:#1}}
\newcommand{\apxref}[1]{Appendix~\ref{apx:#1}}

\renewcommand{\vec}[1]{\begin{bmatrix} #1\end{bmatrix}}

\newcommand{\il}{n} 
\newcommand{\ol}{m} 
\newcommand{\nc}{c} 
\newcommand{\mbt}{M} 
\newcommand{\age}{a} 
\newcommand{\bs}{X} 
\newcommand{\Acal}{\mathcal{A}}
\newcommand{\al}{\alpha} 
\newcommand{\bt}{\beta} 
\newcommand{\gam}{\gamma} 
\newcommand{\lam}{\lambda} 
\newcommand{\titer}{T} 
\newcommand{\prefill}{T_p} 
\newcommand{\itl}{T_{\text{ITL}}} 
\newcommand{\wpref}{W_{\text{pref}}} 
\newcommand{\wdec}{W_{\text{dec}}} 
\newcommand{\wtot}{W_{\text{total}}} 
\newcommand{\ncmin}{c_{\min}}  

\newcommand{\ttft}{T_{\text{TTFT}}}

\newcommand{\Avg}[1]{\overline{#1}}  

\newcommand{\kvu}{u}                       
\newcommand{\ksat}{\kvu_{\text{sat}}}      
\newcommand{\kvlo}{\kvu_{\lo}}             
\newcommand{\kvhi}{\kvu_{\hi}}             
\newcommand{\kvver}{\kvu_{\text{ver}}}     
\newcommand{\thup}{\eta_{\up}}             
\newcommand{\thdn}{\eta_{\dn}}             
\newcommand{\epsshape}{\epsilon_{\text{shape}}} 
\newcommand{\epsrate}{\epsilon_{\text{rate}}}   
\newcommand{\tage}{\tau_{\text{age}}}      
\newcommand{\nver}{N_{\text{ver}}}         
\newcommand{\nwin}{N_{\text{w}}}           
\newcommand{\tload}{\Lambda}               
\newcommand{\kvmax}{K_{\max}}           
\newcommand{\kvreq}{K_{\text{req}}}     
\newcommand{\lo}{\text{lo}}
\newcommand{\hi}{\text{hi}}
\newcommand{\up}{\text{up}}
\newcommand{\dn}{\text{dn}}

\newcommand{\et}{\textit{et al.}}

\newtoggle{hidden}

\begin{document}

\title{An Approximate Queueing Model of LLM Inference Serving
for SLO-Driven Autoscaling}

\iftoggle{hidden}{
	\author{}
}
{
\author{\IEEEauthorblockN{Vishakha Ramani}
\IEEEauthorblockA{\textit{IBM T. J. Watson Research Center} \\
Yorktown Heights, NY \\
vishakharamani@ibm.com}
\and
\IEEEauthorblockN{Asser N. Tantawi}
\IEEEauthorblockA{\textit{IBM T. J. Watson Research Center} \\
Yorktown Heights, NY \\
tantawi@us.ibm.com}}
}

\maketitle

\begin{abstract}
Performance models of LLM servers support both latency evaluation and the design of controllers for autoscaling against service level objectives (SLOs) and for inference optimization.
We model the multiplexed execution of prefill and decode operations with a tractable, approximate queueing model under Markovian assumptions.
Three parameters characterize a model--accelerator pair, namely a baseline per-iteration overhead, a per-token compute cost, and a per-token key-value (KV) cache access cost.
The model combines a mean-value analysis of per-iteration work with a state-dependent Markov chain for batch occupancy to predict mean time to first token (TTFT) and inter-token latency (ITL).
We validate these predictions against measurements and show that the three parameters can be estimated from observed latencies.
Over a grid of input and output lengths and arrival rates spanning light to moderate load, the relative error of the average ITL is about 5\% for Llama-3.1-8B and 8\% for Qwen2.5-14B running on an H100 GPU, and the corresponding TTFT errors are 14\% and 16\%.

We then implement an autoscaling controller that uses the model to adjust inference-server replica counts as the workload changes.
On an OpenShift cluster of H100 GPUs it tracks a fourfold load ramp under both latency
targets, missing one in 7 of 127 control cycles, and its in-loop predictions
carry median errors of at most 5\% for TTFT and 9\% for ITL. A decode-throughput analyzer
from an existing autoscaler, which takes no latency target, misses 27 of 128 cycles under
the same controller and load while provisioning 4\% and 28\% fewer replicas.
\end{abstract}

\begin{IEEEkeywords}
autoscaling, queueing theory, LLM inference, parameter estimation, performance modeling.
\end{IEEEkeywords}

\section{Introduction}\label{sec:introduction}

Large language model (LLM) inference services support latency-sensitive applications such as chatbots, coding assistants, and AI-powered search. Their service level objectives (SLOs) constrain response latency even as request rates fluctuate and each request consumes substantial computation and memory bandwidth.
Providers handle this load by replicating models across accelerators such as GPUs or TPUs. Too many replicas waste expensive capacity, whereas too few create queues resulting in SLO violations. Resource allocation must therefore account for how arrival rate, request size, and the serving policy jointly determine latency.
This is naturally a queueing problem. Requests arrive over time, share compute resources, and wait when demand exceeds immediately available capacity. An analytical model can connect workload and configuration to latency, allowing an orchestrator to choose capacity that balances cost against SLO compliance.

Many autoscalers avoid modeling this relationship explicitly and instead respond
when a measured signal crosses a threshold. Such policies are attractive because
they are simple and require little knowledge of the serving system. Their signals,
however, are only indirect indicators of the latency objectives. The same KV-cache
utilization can correspond to different TTFT and ITL values as prompt lengths,
output lengths, and batch composition change. A threshold on measured latency, in
turn, responds only after the degradation has been observed and does not directly
indicate how much additional capacity is required. An analytical model offers a
different control primitive. It can predict both latency metrics for candidate
replica counts before actuation and allow the controller to select capacity against
explicit SLO constraints.

Our objective is therefore twofold, namely to develop a tractable analytical model of an
LLM inference server and to use that model in an SLO-driven autoscaling controller.
The paper makes three contributions toward this objective.

First, we formulate a three-parameter queueing model for an inference server using
continuous batching (\secref{model} and \secref{analysis}). The model combines a
mean-value analysis of the work performed in each iteration with a state-dependent
Markov chain for the number of active requests. It represents the distinct costs
of prefill computation, decode computation, and KV-cache access, and predicts mean
TTFT and ITL for both unlimited and chunked prefill. This provides the controller
with a direct mapping from workload and serving configuration to the two latency
metrics it is intended to control.

Second, we fit and validate the model on 224 operating points from
vLLM~\cite{vllm,Kwon2023} serving Llama-3.1-8B and Qwen2.5-14B on an H100 GPU
(\secref{validation}). The experiments span 16 pairs of mean input and output
lengths, ranging from 64 to 4096 tokens, and arrival rates from light to moderate
load. The relative error of the average ITL is about 5\% for Llama-3.1-8B and 8\%
for Qwen2.5-14B, while the corresponding TTFT errors are 14\% and 16\%. The fitted
compute and KV-cache parameters also scale consistently with the architectural
differences between the two models, providing a physical check on their
interpretation.

Third, we turn the model into a control mechanism rather than treating it only as
an offline predictor. An initial multi-observation fit and an online sliding-window
estimator calibrate the model from observed latencies as operating conditions
change (\secref{estimation}). A queueing-model analyzer then evaluates candidate
replica counts against the TTFT and ITL targets, and an actuator applies the chosen
allocation (\secref{control-loop}). We deploy this controller in an OpenShift
cluster of H100-backed vLLM servers and compare it against a decode-throughput
analyzer, holding the surrounding control loop and workload fixed
(\secref{baseline}). That analyzer provisions enough replicas to sustain the
decode-token demand at a target KV-cache utilization, and it takes no latency target
at all. The comparison exposes a capacity--latency tradeoff. Our controller misses a
latency target in $7$ of $127$ control cycles against $27$ of $128$ for the
throughput analyzer, which in exchange provisions fewer replicas on average. We discuss the implications of these
results and directions for future work in \secref{discussion}.

\section{Background}
LLM inference is an autoregressive process that transforms an input prompt into an output sequence.
Inference begins with tokenization, which decomposes the raw input text into tokens.
This preprocessing is inexpensive relative to the subsequent model computation.
Processing a request consists of repeatedly running a \emph{forward pass}, a single execution of the model's transformer layers that maps the current per-token hidden states to per-token outputs.
Each layer is dominated by two costs, namely dense matrix multiplications between hidden states and the model weights, and an \emph{attention} computation that updates each token's representation using a weighted combination of the key and value (KV) vectors of all preceding tokens in its sequence.
This processing happens in two serial phases that have very different computational profiles.

In the first phase, called \emph{prefill}, the server runs a single forward pass over all input tokens of the request and stores their KV vectors in a per-request \emph{KV cache} so that subsequent steps need not recompute them. The final hidden state of the prompt then seeds the decode phase, whose first step produces the first output token.
Prefill is compute bound because all input tokens are processed in parallel and the cost of the dense linear operations grows with the input length.
In the second phase, called \emph{decode}, the server generates the output tokens one at a time, where each step runs a forward pass on a single new token, appends its KV vectors to the cache, and lets the attention computation read the entire cache to predict the next token.
A single decode step is memory bound because it generates only one new token, whose compute cost is small, while the attention computation must read the full KV cache from high-bandwidth memory (HBM), and the cost of this read grows with the request's sequence length.

To increase GPU utilization, modern inference servers run a single forward pass over tokens from multiple requests concurrently.
A \emph{batch} is the set of tokens that the server schedules together in one forward pass, and the execution of that pass is called an \emph{iteration}.
Depending on the scheduling discipline, a batch may consist of input tokens from a single prefilling request, input tokens from several prefilling requests, one decode token from each of several decoding requests, or a mix of input and decode tokens drawn from both prefilling and decoding requests.
Regardless of the mix, the model weights are loaded from HBM once per layer per iteration and the matrix multiplications that consume them amortize the load over every token in the batch.
The per-token compute cost and per-request KV-cache read cost still grow with the batch token count and each request's sequence length, respectively, so larger batches and longer sequences both increase iteration time.

Modern serving systems adopt \emph{continuous batching}~\cite{Yu2022}, where requests do not wait to be assembled into a synchronized batch and do not wait for the other requests in their batch to finish.
Instead, on every iteration boundary, a request that has just completed its decode phase departs and any waiting request can be admitted into the batch and start its prefill in the very next iteration.
The batch composition therefore varies continuously over time, with requests at different phases coexisting in the same iteration.

In the following section, we develop an analytical model for latency in an LLM inference server employing continuous batching.
We represent each request by the work it adds to the system and model iteration
time as the total work of all active requests.

\section{Related Work}\label{sec:related}

\paragraph{Serving mechanisms and benchmarking}
Prior systems work introduced the scheduling mechanisms represented by our model.
Yu \et~\cite{Yu2022} introduce continuous batching, where requests join and leave the active batch at iteration boundaries rather than waiting for a full batch to complete.
Agrawal \et~\cite{Agrawal2024} introduce chunked prefill, splitting a long input prompt into smaller chunks that are interleaved with ongoing decode steps, and evaluate the resulting throughput-latency tradeoff against other scheduling disciplines.
Patel \et~\cite{Patel2024} propose a method for splitting the prefill and decode phases and evaluate its performance.
A complementary line of work characterizes servers empirically rather than analytically.
Lazuka \et~\cite{Lazuka2024} perform extensive benchmarking and build a system for characterizing and predicting performance.
Such black-box predictors can be accurate within their training configurations, but each new model--accelerator pair requires another measurement campaign. Our three-parameter formulation is intended to reduce that calibration burden.

\paragraph{Queueing models of inference serving}
Mitzenmacher and Shahout~\cite{Mitzenmacher2025} provide an overview of the challenges and open problems in LLM performance modeling and analysis.
Yang \et~\cite{Yang2024} use a queueing model to evaluate LLM request latency, and Ao \et~\cite{Ao2025} develop a fluid approximation of LLM serving and introduce a scheduling algorithm with near-optimal performance.
Closest to our work is Inoue~\cite{Inoue2021}, who gives a closed-form characterization of a GPU-based inference server under \emph{dynamic} batching, where a batch is assembled, executed, and completed as a unit, with a batch-size-dependent processing time.
Our setting differs in three ways that require a different analysis, not just different parameters.
First, under continuous batching a request joins and departs at iteration boundaries, so the batch composition is a mixture of requests at different points in their lifetime instead of a set of jobs that enter and leave together.
Second, an autoregressive request remains active through one or more prefill
iterations and then through a separate decode iteration for each generated token.
These two phases have very different computational costs, so the request cannot be
represented as a single processing step.
Third, the cost of each iteration grows with the number of prompt and generated
tokens stored in the active requests' KV caches. Prompt length and generated
response length therefore affect TTFT and ITL differently.
Consequently our service time is state-dependent through both the occupancy and the token budget, and the model predicts two latency metrics rather than one.

\paragraph{Model-driven autoscaling}
Using a performance model to drive resource allocation has a long history outside LLM serving.
Gandhi \et~\cite{Gandhi2014} build an adaptive, model-driven autoscaler for cloud applications that refines a queueing-model estimate from live measurements. Our control loop adopts a similar separation between model analysis and online tuning.
Gujarati \et~\cite{Gujarati2017} target machine learning inference specifically, using a model of per-request demand to autoscale while meeting service level agreements with high resource efficiency.
Both predate the two-phase autoregressive workload and KV-cache pressure of LLM serving, which our model explicitly represents.
Among recent autoscalers and serving systems for LLM inference, related goals include GPU heterogeneity for cost~\cite{Griggs2024}, joint placement and scaling for SLO compliance~\cite{Nie2024,Turkkan2024}, mixed-SLO and multi-class workloads~\cite{Zhang2025,Chen2025,Zhu2025}, prefill-decode multiplexing~\cite{Cui2025}, and energy efficiency~\cite{Stojkovic2024}.
Most directly related is SageServe~\cite{Jaiswal2025}, which combines hourly forecasts of per-model regional token demand with benchmarked model--accelerator throughput capacities in an integer linear program that coordinates long-timescale model placement and request routing across data centers.
It complements this planner with utilization-triggered scaling, shared capacity for interactive and non-interactive workloads, and deadline-aware request scheduling.
SageServe therefore addresses where model capacity should be placed across regions ahead of forecast demand, whereas our work addresses how many replicas a particular serving configuration requires to satisfy explicit TTFT and ITL constraints.
Its online allocation formulation operates on forecasted throughput capacity. Our analyzer instead models continuous batching, prefill and decode work, and KV-cache access to predict both latency metrics, with its three parameters continuously re-estimated from observed latencies.
The approaches are complementary. SageServe-style forecasts could supply demand one provisioning delay ahead to our analyzer, while our queueing model could provide latency-aware capacity constraints to a global placement planner.
The same self-tuning analyzer can therefore support different models and accelerators without training a separate black-box latency predictor for each deployment.

\section{System Model} \label{sec:model}
Consider an LLM server whose requests arrive as a Poisson process with rate $\lambda$.
This assumption is what reduces the occupancy process to a birth--death chain, and it is
therefore what makes the resulting latency expressions closed form. It is also the arrival
model used by the closed-form queueing analyses of inference serving reviewed in
\secref{related}.
Measured production traffic is more variable than this. A 213-day trace of Azure OpenAI
services fits its request arrivals with a Gamma interarrival distribution whose fitted
shape, and therefore whose coefficient of variation, changes sharply from one 20-minute
window to the next, and reports higher arrival burstiness than a Microsoft Azure Functions
trace at the same mean request rate~\cite{Wang2024}. A Poisson process holds its
coefficient of variation fixed at one, so it represents neither that level of variability
nor its drift over time. At a fixed mean rate our waiting-time and TTFT predictions are
therefore optimistic under bursty arrivals, and increasingly so as load rises toward
saturation.

Nevertheless, Poisson arrivals are what let the model act as a control primitive, for
three reasons. First, the only arrival statistic a controller can measure cheaply and
continuously is the mean rate, so a richer arrival model would carry parameters that no
routine measurement identifies. Second, representing correlated arrivals would either
enlarge the state beyond the single occupancy variable, as an arrival process modulated by
its own Markov chain does, or give up the closed-form occupancy distribution for a numerical
matrix-analytic solution. Either way the model must be solved afresh for every candidate
replica count a controller considers, and the closed form is what keeps that affordable. Third, the sign of the resulting error is known rather than
arbitrary, since positive arrival correlation raises queueing delay, so an operator can
absorb it in the SLO margin.

Re-estimating the rate limits the exposure further. Our controller refreshes $\lambda$
from measured arrivals at regular intervals, so demand variation on timescales longer than
one interval, including diurnal variation, is tracked by re-estimation rather than left to
the arrival model. What the model does not capture is arrival correlation within a single
interval, and this is the case for eventually making arrival variability an explicit model
input.

Each request has an input length $\il>0$ and an output length $\ol>0$, which we assume are mutually independent.
Throughout the analysis, we set $\il$ and $\ol$ to the workload's mean input and output lengths. The resulting homogeneous-batch approximation assigns every request the work of an average request.

During its lifetime in the server, an $(\il,\ol)$ request first waits in a queue, then passes through $\il$ prefill-token stages and $\ol$ decode-token stages. A decode iteration advances the request by one stage, whereas a prefill iteration may process any number of stages in $\{1,\ldots,\il\}$, depending on the scheduling discipline.
The request therefore waits until it is admitted, completes prefill in one or more iterations, and then completes exactly $\ol$ decode iterations.

Our analysis uses three model--accelerator-specific parameters that convert work to wall-clock time, namely $\al$, $\bt$, and $\gam$.
Parameter $\al$ represents the baseline time per iteration irrespective of batch composition, accounting for kernel launch latencies, synchronization barriers, and weight-loading overhead.
Parameter $\bt$ is the compute time per token during the forward pass, governed by the accelerator's peak floating-point throughput, and $\gam$ is the KV-cache memory access time per token, governed by its memory bandwidth.
Since all three parameters carry units of time, work quantities derived from them are expressed directly in units of time, consistent with the standard queueing-theoretic notion of workload as service time.

We consider the general case where a maximum token budget limits the
total number of tokens processed per iteration. A request with input length
$\il$ tokens and output length $\ol$ tokens is represented by
$\nc \geq 1$ equal-sized effective prefill stages of $\il/\nc$
tokens~\cite{Agrawal2024}, followed by $\ol$ decode stages. This representation
approximates the latency-smoothing effect of chunked prefill without treating
$\nc$ as the literal number of chunks selected by the runtime scheduler.
We derive the occupancy-dependent mean-field approximation $\nc(i)$ in
\apxref{chunks}.
With unlimited token capacity, prefill completes in a single iteration. This is the special case $\nc = 1$, as shown in
\apxref{unlimited}.

\section{Analysis}\label{sec:analysis}
\subsection{Per-iteration work model}\label{sec:analysis:iteration}

We characterize the state of a request during a given iteration by its
\emph{processing age} $\age \in \Acal$, with state space
\begin{equation*}
    \Acal = \{0_1, 0_2, \ldots, 0_\nc, 1, 2, \ldots, \ol\},
\end{equation*}
where $0_j$ denotes that the $j$-th prefill chunk is being processed and
$k \in \{1,\ldots,\ol\}$ denotes that the $k$-th decode step is in progress.
The request departs the system upon completing state $\ol$.
Within the effective-stage approximation, a request advances by one stage per
iteration and spends one effective iteration in each of its $\nc+\ol$ states.
Across its effective iterations, the request is therefore equally likely to be
in any state. The probability $p_a$ that an active request is in state $\age$ is
$p_a=1/(\nc+\ol)$ for every $\age\in\Acal$. The work model below uses this
per-iteration distribution to average a request's contribution.

The work contributed by a request in a given iteration depends on its current
state. During prefill chunk $j$, for $j = 1, \ldots, \nc$, the system computes
the KV vectors of $\il/\nc$ tokens at compute cost $\bt\il/\nc$ and writes them
to the KV cache at memory cost $\gam\il/\nc$. Computing attention scores
additionally requires reading the KV vectors of the $(j-1)\il/\nc$ tokens
already processed in prior chunks, at cost $\gam(j-1)\il/\nc$, so the total work
for prefill chunk $j$ is $W_{0_j} = \bt\il/\nc + \gam j\il/\nc$. The memory
term $\gam j\il/\nc$ grows with $j$, since early chunks attend over a small KV
cache while later chunks read all previously cached tokens, making later
chunks progressively more memory-bound. Summing over all $\nc$ chunks gives the
total prefill work
\begin{equation}\eqnlabel{wpref}
    \wpref = \sum_{j=1}^{\nc} W_{0_j} = \bt\il + \gam\il\frac{\nc+1}{2}.
\end{equation}
Throughout the analysis we reserve $j$ for the prefill chunk index, $k$ for the
decode step index, and $i$ for the system occupancy state of the Markov chain.

During decode step $k$, for $k = 1, \ldots, \ol$, the system computes the KV
vectors of the new output token at cost $\bt$, reads the KV cache holding all
$\il$ prefill tokens plus the $k-1$ tokens generated so far at cost
$\gam(\il+k-1)$, and writes the new token's KV vectors to the cache at cost
$\gam$. The total work for decode step $k$ is therefore
$W_k = \bt + \gam(\il + k)$. Summing over all $\ol$ decode steps gives
\begin{equation}\eqnlabel{wdec}
    \wdec = \sum_{k=1}^{\ol} W_k = \bt\ol + \gam\ol\!\left(\il + \frac{\ol+1}{2}\right).
\end{equation}
The decode work scales with the output length $\ol$. Each token incurs compute
cost $\bt$ and an increasing memory cost as the KV cache grows.

Summing \eqnref{wpref} and \eqnref{wdec}, the total work over the request's lifetime is
\begin{align}
\eqnlabel{wtotal}
\wtot &= \wpref + \wdec \nn
&= \bt(\il+\ol) + \gam\paren{\frac{\il(\nc+1)}{2} + \ol\paren{\il + \frac{\ol+1}{2}}}.
\end{align}
The compute term $\bt(\il+\ol)$ reflects the cost of processing all $\il+\ol$
tokens in the forward pass. The memory term has two parts. The first part,
$\gam\il(\nc+1)/2$, is the attention read cost accumulated across all prefill
chunks, growing with $\nc$, and the second part, $\gam\ol(\il+(\ol+1)/2)$, is the
attention read cost across all decode steps, growing with both $\il$ and $\ol$.

Let $\delta$ be the marginal cost of adding a concurrent request to the batch.
Under the assumption that each of the $\nc+\ol$ states is equally likely, this cost is
\begin{align}\eqnlabel{delta}
    \delta &= \frac{\wtot}{\nc+\ol}.
\end{align}
When the batch holds $\bs$ active requests, each contributes mean work $\delta$
per iteration, and the iteration time is
\begin{equation}\eqnlabel{titer}
    \titer(\bs) = \al + \bs\delta.
\end{equation}
This linearity has been observed experimentally~\cite{Lazuka2024,Ao2025,Yang2024}.
\subsection{State-dependent Markovian analysis}\label{sec:analysis:markov}

We now derive two latency metrics, namely time to first token, $\ttft$, from request arrival
to the first output token, and inter-token latency, $\itl$, the mean interval between
successive output tokens.
Define prefill latency, $\prefill$, as the time from admission to the batch until completion of the prefill phase.
We treat the prefill phase as building the KV cache, with the first decode step producing the first output token.
Thus, $\ttft$ is the sum of the time a request waits in the queue, its prefill time $\prefill$, and one $\itl$ for the first token.

\begin{figure}[t]
\centering
\resizebox{\columnwidth}{!}{%
\begin{tikzpicture}[
  >=stealth, node distance=11mm,
  state/.style={circle, draw, minimum size=8mm, inner sep=0pt},
  rate/.style={font=\small}]
  \node[state] (s0) {$0$};
  \node[state, right=of s0] (s1) {$1$};
  \node[state, right=of s1] (s2) {$2$};
  \node[right=of s2] (d1) {$\cdots$};
  \node[state, right=of d1] (sB) {$B$};
  \node[state, right=of sB] (sB1) {$B{+}1$};
  \node[right=of sB1] (d2) {$\cdots$};
  \foreach \a/\b in {s0/s1, s1/s2, s2/d1, d1/sB, sB/sB1, sB1/d2}
    \draw[->] (\a) to[bend left=40] node[rate, above]{$\lambda$} (\b);
  \draw[->] (s1)  to[bend left=40] node[rate, below]{$\mu(1)$}   (s0);
  \draw[->] (s2)  to[bend left=40] node[rate, below]{$2\mu(2)$}  (s1);
  \draw[->] (d1)  to[bend left=40] (s2);
  \draw[->] (sB)  to[bend left=40] node[rate, below]{$B\mu(B)$}  (d1);
  \draw[->] (sB1) to[bend left=40] node[rate, below]{$B\mu(B)$}  (sB);
  \draw[->] (d2)  to[bend left=40] node[rate, below]{$B\mu(B)$}  (sB1);
\end{tikzpicture}}
\caption{Birth--death Markov chain over the number of requests $i$ in the system. Arrivals occur at rate $\lambda$ and raise the state by one. With $\mu(i)$ the per-request departure rate in state $i$, the total departure rate is $i\,\mu(i)$ while the batch is below the maximum size $B$, saturating at $B\,\mu(B)$ for $i > B$.}
\label{fig:birth-death}
\end{figure}
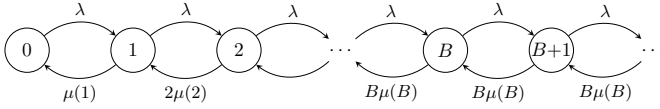

\paragraph{The queueing model}
We model the server as a birth--death Markov chain (\Figref{birth-death}) whose state $i \ge 0$ is the
number of requests in the system. Up to $B$ requests are served together as a
batch, where $B$ is the maximum batch size, and the rest wait in a queue.
Arrivals produce upward transitions at rate $\lambda$ in every state. A request in
a batch of $i$ has service time $\tau(i)$ and therefore completes at rate
$\mu(i) = 1/\tau(i)$. With $i$ requests present, the down-transition rate is
$i\,\mu(i)$, saturating at $B\,\mu(B)$ for $i > B$. The rate is load-dependent,
so the chain is state-dependent. We derive $\tau(i)$ next.

\paragraph{Prefill and inter-token latency}
A request completes $\nc$ prefill iterations followed by $\ol$ decode iterations. We follow
one such request through the system and call it the tagged request.
Each iteration includes the tagged request's work for its current stage and the
baseline plus the work of the other $i-1$ requests, represented by $\titer(i-1)$.
Summing the tagged request's work over its $\nc$ prefill iterations gives
\begin{align}\eqnlabel{prefill-i}
\prefill(i) &= \nc\,\titer(i-1) + \wpref. 
\end{align}
During decode, the inter-token latency is the iteration time due to the other
$i-1$ requests plus the tagged request's mean decode-step work
\begin{align}\eqnlabel{itl-i}
\itl(i) &= \titer(i-1) + \frac{1}{\ol}\sum_{k=1}^{\ol} W_k. 
\end{align}
\paragraph{Service time}
A request's service time is its prefill latency plus its $\ol$ inter-token gaps,
\begin{equation}\eqnlabel{tau-def}
\tau(i) = \prefill(i) + \ol\,\itl(i), \qquad 1 \le i \le B.
\end{equation}
Substituting equations \eqnref{prefill-i} and \eqnref{itl-i} into \eqnref{tau-def} and using \eqnref{wtotal} gives
\begin{align}\eqnlabel{tau}
\tau(i) &= (\nc+\ol)\,\titer(i-1) + \wtot \nn
        &= (\nc+\ol)\,\titer(i), \qquad 1 \le i \le B.
\end{align} 
The second line follows from two facts. First, $\titer(i-1) = \titer(i) - \delta$,
so each of the $\nc+\ol$ iterations was charged at $\delta$ below the full
iteration time $\titer(i)$. Second, the request's total work is
$\wtot = (\nc+\ol)\delta$, which is exactly $\delta$ for each of those $\nc+\ol$
iterations. Adding $\wtot$ therefore restores the missing $\delta$ to every
iteration, turning each $\titer(i-1)$ into a full $\titer(i)$, so the service time
is simply $\nc+\ol$ iterations at batch size $i$.

The effective prefill-stage count is also state-dependent. Because the token
budget couples it to the batch size, we evaluate $\nc=\nc(i)$ from
\apxref{chunks} at each state $i$ and
compute $\wpref$, $\wtot$, $\delta$, $\titer(i)$, and $\tau(i)=(\nc(i)+\ol)\titer(i)$
at that state's $\nc(i)$, so the telescoping above holds within each state. The
chain stays one-dimensional because $\nc(i)$ is a deterministic function of the
occupancy $i$ rather than an added state variable.

\paragraph{Steady state}
Let $\pi_i$ be the steady-state probability of finding $i$ requests in the system.
Writing
\begin{equation*}
g(i) = \frac{\lambda}{i\,\mu(i)}, \qquad i = 1, 2, \ldots, B,
\end{equation*}
the balance equations give
\begin{equation*}
\pi_i = \left\{
\begin{array}{ll}
\pi_0 \prod_{l=1}^{i} g(l), & i = 1, 2, \ldots, B, \\[4pt]
\pi_0 \left(\prod_{l=1}^{B} g(l)\right) g(B)^{\,i-B}, & i > B,
\end{array}
\right.
\end{equation*}
with $\pi_0$ fixed by the normalization $\sum_{i=0}^\infty \pi_i = 1$.
When all $B$ batch slots are occupied, the server completes requests at rate
$B/\tau(B)$. Writing the offered load as
\begin{equation}\eqnlabel{rho}
\rho = \frac{\lambda\,\tau(B)}{B},
\end{equation}
a steady state therefore exists only when $\rho<1$, equivalently
$\lambda<B/\tau(B)$. All steady-state quantities below assume this condition,
and we refer to $\rho$ as the utilization throughout.

\paragraph{Latency metrics}
Requests are queued only when more than $B$ are present, which happens with probability $P_w = \sum_{i>B} \pi_i$.
We summarize the chain by the mean number of requests in the system $\Avg{N}$,
the mean number in service (or mean batch size) $\Avg{X}$, and the mean number
waiting $\Avg{Q}$
\begin{align*}
\Avg{N} &= \sum_{i=0}^\infty i\,\pi_i, \nn
\Avg{X} &= \sum_{i=0}^{B} i\,\pi_i + B\,P_w, \nn
\Avg{Q} &= \Avg{N} - \Avg{X}.
\end{align*}
Here $\Avg{X}$ is the steady-state mean of the batch occupancy $i$ used above, so it plays the role of the working batch size. The mean $\Avg{N}$ adds the requests that are merely waiting.
By Little's law the mean service time and mean waiting time are $\Avg{S} = \Avg{X}/\lambda$ and $\Avg{W} = \Avg{Q}/\lambda$, respectively.
The mean batch size $\Avg{X}$ also gives the average iteration time,
\begin{equation*}
\titer(\Avg{X}) = \al + \Avg{X}\delta.
\end{equation*}
From $\tau(i) = (\nc+\ol)\titer(i)$ and $\Avg{S} = \Avg{X}/\lambda$, the mean
service time obeys $\Avg{X} \approx \lambda(\nc+\ol)\titer(\Avg{X})$, which solves
to $\titer(\Avg{X}) \approx \al/(1-\lambda(\nc+\ol)\delta)$ when the batch is
unsaturated. Its denominator requires
$\lambda(\nc+\ol)\delta = \lambda\wtot < 1$.

Because $\prefill(i)$ is linear in $i$ at a fixed chunk count, the mean
prefill latency $\sum_{i=0}^{B} \pi_i\,\prefill(i) + P_w\,\prefill(B)$ is
approximated by $\prefill(\Avg{X})$. Evaluating the metrics at the mean batch size
$\Avg{X}$ and its chunk count $\nc(\Avg{X})$, we obtain
\begin{equation*}
\itl(\Avg{X}) = \frac{\Avg{S} - \prefill(\Avg{X})}{\ol},
\end{equation*}
and
\begin{equation*}
\ttft = \Avg{W} + \prefill(\Avg{X}) + \itl(\Avg{X}).
\end{equation*}
We read $\itl$ off the mean service time as $(\Avg{S} - \prefill(\Avg{X}))/\ol$ rather than
from equation \eqnref{itl-i} directly. The two agree when the batch is rarely full, and the
former also captures how queueing changes the effective batch size.

\section{Experimental Validation}\label{sec:validation}

To validate the model, we ran separate experiments on a vLLM~\cite{vllm,Kwon2023}
server with an H100 GPU, serving Llama-3.1-8B~\cite{llama3} in one experiment and
Qwen2.5-14B~\cite{qwen25} in the other. GuideLLM~\cite{guidellm} generated the traffic.
We placed the GuideLLM generator on the same node as the vLLM server to keep network jitter out of the latency measurements.
For each experiment, we swept a $4 \times 4$ grid of mean input and output lengths
drawn from $\{64,256,1024,4096\}$, giving 16 length pairs. Within each sweep, every
request drew its input and output lengths independently and uniformly from
$[x/2,3x/2]$ around the corresponding mean $x$.
For each sweep, GuideLLM ran a synchronous case (one request at a time), a
throughput-saturating case, and eight Poisson-arrival runs with rates spaced between
these two extremes.
We excluded the throughput-saturating run and the two highest-rate Poisson runs,
which approach saturation and require very long traces for stable statistics. This
left seven runs per sweep and 112 data points per model.
Because the model takes scalar input and output lengths, we evaluated it at each
run's mean input and output token counts.
Individual lengths therefore vary by a factor of three within a run, with a coefficient of
variation near $0.29$, so the accuracy reported below is obtained under real length
dispersion rather than with every request identical. That dispersion is what the
homogeneous-batch approximation replaces with a single mean request.
We evaluated the queueing model with a per-iteration token budget $\mbt = 8192$,
which sets the chunk count $\nc$, and a maximum batch size $B = 256$.
The value $B=256$ is the vLLM default for \texttt{max\_num\_seqs}. We impose no
smaller administrative cap, allowing the validation to exercise the model's
state-dependent region.

Because the excluded runs are the ones nearest saturation, the accuracy figures reported
below characterize light to moderate load rather than the regime $\rho \to 1$. This is the
regime the model exists to serve. Predicted latency diverges as $\rho \to 1$, so any
capacity decision that holds TTFT and ITL to finite targets keeps the server away from
saturation, and the steady-state chain is undefined once $\rho \geq 1$ in any case.
Extending the validation toward saturation with longer traces would locate the accuracy
boundary more precisely, and we leave that to future work.

\begin{table}[t]
	\centering
	\footnotesize
\setlength{\tabcolsep}{4pt}
\begin{tabular}{l|ccc|cc}
\hline
Model & $\alpha$ & $\beta$ & $\gamma$ & ITL err & TTFT err \\
\hline
Llama-3.1-8B  & $6.68$ & $2.01{\times}10^{-2}$ & $5.52{\times}10^{-5}$ & $4.6\%$ & $13.6\%$ \\
Qwen2.5-14B   & $10.14$ & $3.68{\times}10^{-2}$ & $8.48{\times}10^{-5}$ & $7.9\%$ & $15.8\%$ \\
\hline
\end{tabular}

	\vspace{5pt}
	\caption{Joint fit of $\{\alpha,\beta,\gamma\}$ in milliseconds on a 16-sweep grid for two models on an H100 GPU.}
	\label{tab:validation-fit}
\end{table}

A Nelder-Mead optimizer fits the parameters $\theta=\{ \alpha, \beta, \gamma \}$ that minimize the per-point sum of squared relative errors in the predicted TTFT and ITL against the measurements.
The relative form is scale-free, so the two metrics enter the loss on equal footing despite TTFT being roughly an order of magnitude larger than ITL in absolute terms.
We start from $\theta_0=\{1, 10^{-2}, 10^{-4}\}$, scaling each parameter by its starting value to keep the simplex well-conditioned across the orders of magnitude that separate them.

The fitted parameters and the relative errors are reported in \tabref{validation-fit}, where each relative error is the mean absolute deviation divided by the mean measurement.
The fitted $\beta$ and $\gamma$ are larger for Qwen2.5-14B than for Llama-3.1-8B,
consistent with their interpretation as per-token compute and KV-cache access times.
Parameter $\beta$ should track the per-token compute of a forward pass, which is dominated by matrix multiplies through the model weights and therefore scales roughly with the parameter count.
The two models have $8$B and $14$B parameters, a ratio of $1.75$, and the fitted $\beta$ rises from $2.01 \times 10^{-2}$ to $3.68 \times 10^{-2}$, a ratio of $1.83$.
Parameter $\gamma$ should track the per-token KV-cache access time, which is set by the KV-cache size that the attention kernel reads at each step.
The KV-cache size per token is $2 \times L \times H \times d_h$, where $L$ is the number of transformer layers, $H$ is the number of key-value heads, and $d_h$ is the head dimension.
The two models share the same number of KV heads ($H=8$) and the same head dimension ($d_h=128$) but have $L=32$ and $L=48$ layers, respectively, a ratio of $1.5$.
The fitted $\gamma$ rises from $5.52 \times 10^{-5}$ to $8.48 \times 10^{-5}$, a ratio of $1.54$, in line with this architectural difference.

\begin{figure}[t]
	\centering
	\includegraphics[width=1.0\columnwidth]{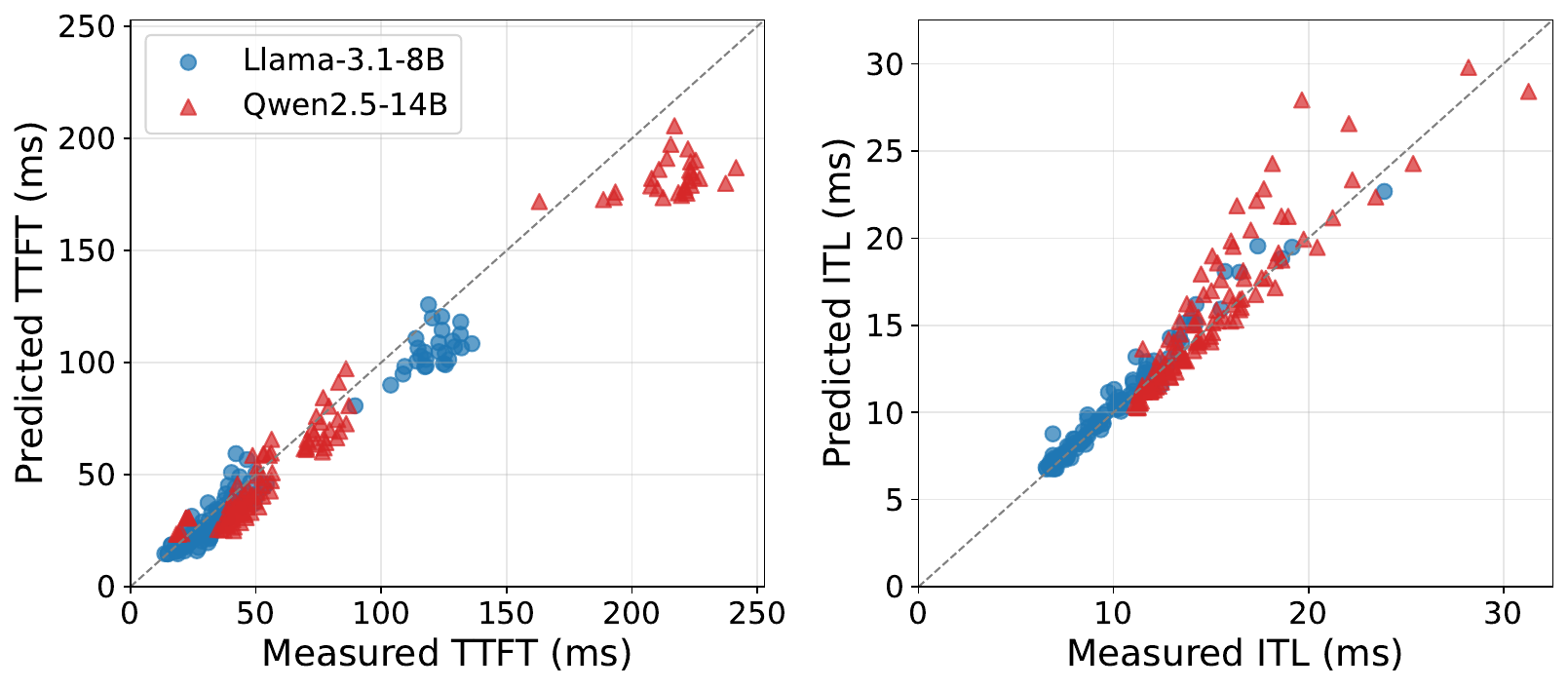}
	\caption{Predicted vs.\ measured TTFT (left) and ITL (right) for Llama-3.1-8B (blue circles) and Qwen2.5-14B (red triangles). The dashed line is $y=x$.}
	\label{fig:validation-scatter}
\end{figure}

We checked the physical plausibility of these values against H100 specifications.
For Llama-3.1-8B, each token in the forward pass costs about $2 \times 8 \times 10^9 = 16$ GFLOPs of work.
This assumes that a forward pass costs roughly twice the parameter count in floating-point operations~\cite{kaplan2020scalinglawsneurallanguage}.
The H100 has a peak BF16 throughput near $989$ TFLOPS.
The resulting ideal compute time is about $1.6 \times 10^{-2}$ ms per token.
The fitted $\beta = 2.01 \times 10^{-2}$ ms corresponds to about 80\% of peak compute throughput.
A similar check applies to $\gamma$.
The KV cache stores $2 L H d_h$ values per token, which for Llama-3.1-8B in BF16 is about $128$ KB per token.
The H100 HBM bandwidth is roughly $3.35$ TB/s.
Reading one cached token at peak bandwidth therefore takes about $3.9 \times 10^{-5}$ ms.
The fitted $\gamma = 5.52 \times 10^{-5}$ ms corresponds to about 71\% of peak bandwidth.
This is a check the fit could have failed. The parameters come from latency measurements
alone, with no knowledge of the accelerator's floating-point throughput or memory
bandwidth, so nothing in the procedure prevents a three-parameter fit from returning a
per-token time faster than the hardware allows, which would leave $\bt$ and $\gam$ without
the physical reading the model assigns them. Both instead land on the slow side of their
bounds, by margins appropriate to the operation each one represents.

For $\gam$ that operation is a streaming read. A microbenchmark attains 90\% and 91\% of
nominal peak memory bandwidth on A100 and H800 parts~\cite{Luo2024}, so the 71\% implied by
$\gam$ sits below the practical ceiling, which is where an attention kernel belongs
when it reads a KV cache held in noncontiguous blocks rather than streaming contiguous
memory. For $\bt$ that operation is a dense matrix multiplication, since $\bt$ tracks the
per-token cost of the weight matrix multiplications. A tuned kernel for that operation
reaches 80\% to 90\% of peak device throughput while an optimized attention kernel reaches
only 50\% to 73\%~\cite{Dao2023}, and the 80\% implied by $\bt$ sits at the base of that
range.

Two effects place it at the base of that range rather than the middle, and both work in the
same direction. First, the ideal time assumes a forward pass costs twice the parameter
count in floating-point operations, which omits attention and normalization work and so
understates the work each token requires. Second, $\bt$ is a fitted aggregate rather than a
measured kernel time, so it absorbs per-token costs that the model does not represent
separately. Each effect lowers the computed fraction, which makes 80\% a lower bound on the
achieved matrix-multiply efficiency rather than a point estimate.

We read these figures as evidence that $\bt$ and $\gam$ carry the physical meaning the
model assigns them, not as measurements of kernel efficiency.
A rigorous cross-check against kernel-level profiling is left to future work.

\Figref{validation-scatter} compares predicted and measured TTFT and ITL across all $224$ points.
The points track $y=x$ closely over two orders of magnitude in latency. The largest
deviations occur for Qwen2.5-14B in the longest-sequence, highest-load sweeps, where
the mean-field approximation is least accurate.

\section{Parameter Estimation}\label{sec:estimation}

The three model parameters $\alpha$, $\beta$, and $\gamma$ depend on the model and accelerator pair and on operational conditions that vary at runtime, such as KV-cache pressure and request batching.
We therefore estimate them online with a \emph{model tuner}, the component that keeps the
queueing model calibrated from observed operating points as load and traffic mix evolve.
Its input arrives on a repeating \emph{control cycle}, which records, for each managed
deployment, the arrival rate $\lam$, the average input and output token counts, and the
measured TTFT and ITL, producing one observation per cycle.
The tuner uses an initial fit followed by an online sliding-window estimator.

On first use of a (model, accelerator) pair, the tuner accumulates a configurable number of initial observations across consecutive control cycles and then runs a Nelder-Mead fit to find the parameter vector $\theta = \{\alpha, \beta, \gamma\}$ that jointly minimizes the mean squared relative error in the predicted TTFT and ITL against the recorded observations.
Variables are scaled by the starting point so the simplex is well-conditioned across parameters that span several orders of magnitude. 
We use multiple observations rather than solving for the three parameters from a single
observation treated as queue-free. At moderate to high utilization the light-traffic
assumption behind that shortcut breaks down and inflates $\alpha$, because queueing delay
present in the measurement is charged to the per-iteration overhead. The multi-observation
fit does not require the assumption.

After the initial fit, the tuner refreshes its parameter estimate at every control cycle by re-running the same Nelder-Mead optimization on the most recent $\nwin$ observations, where $\nwin$ is a configurable window size.
The window provides enough operating points for a well-conditioned fit while letting the estimate respond to drift in $\alpha$, $\beta$, or $\gamma$ caused by changing batch composition, KV-cache occupancy, or background tenancy on the accelerator.
The previous parameter vector initializes the next simplex, reducing the per-cycle
optimization cost and encouraging continuity across cycles.
Observations from saturated cycles are excluded because the steady-state queueing
model is undefined when $\rho \geq 1$.

\section{Model-Based Autoscaling Control Loop}\label{sec:control-loop}

\subsection{Description of model-based autoscaler}\label{sec:control-loop:controller}

\begin{figure}[t]
	\centering
	\includegraphics[width=1.0\columnwidth]{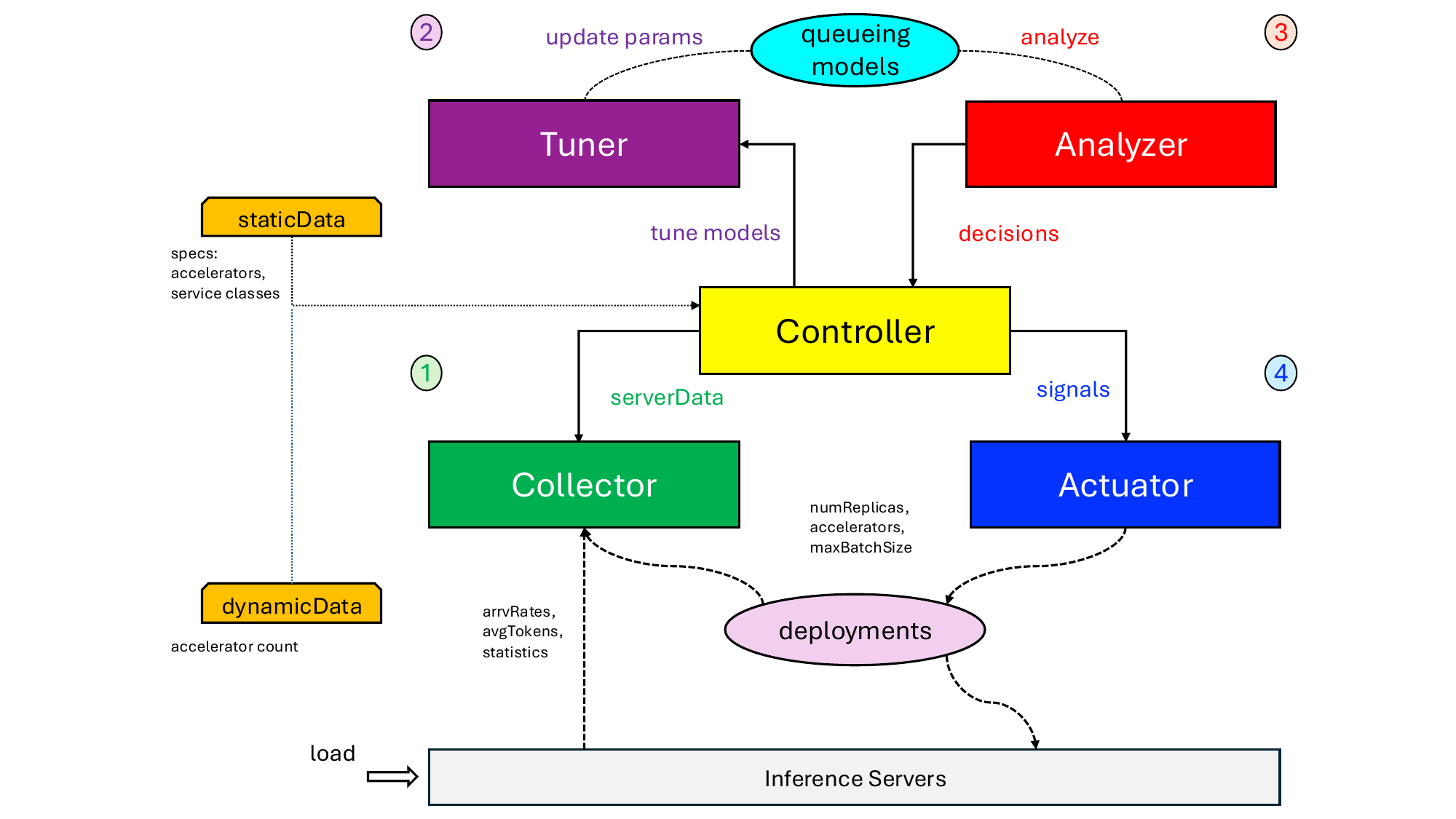}
	\caption{Autoscaling control loop. Colored rectangles are the five microservices, orange parallelograms are configuration inputs, and ovals are shared state (queueing models and Kubernetes deployments). The numbered labels mark the four steps of one control cycle, namely collect (1), tune (2), optimize (3), and actuate (4).}
	\label{fig:control-loop}
\end{figure}

We embed the queueing model and parameter tuner in a closed-loop autoscaler that
adjusts inference-deployment replica counts to meet ITL and TTFT SLOs at minimum cost.
The autoscaler is part of the llm-inferno framework~\cite{llm-inferno,llm-d}.
\Figref{control-loop} shows its architecture.
The framework consists of five cooperating microservices, namely the \emph{controller},
\emph{collector}, \emph{tuner}, \emph{analyzer (optimizer)}, and \emph{actuator}. They share
the queueing models and act on the inference deployments.
Each control period has four steps. First, the collector obtains each server's
arrival rate, average input and output token counts, and observed TTFT and ITL, and
returns these measurements as \emph{serverData}. Second, the tuner refreshes
$\theta=\{\alpha,\beta,\gamma\}$ for every model--accelerator pair using the
techniques of \secref{estimation}. Third, the analyzer evaluates candidate replica
counts and accelerator assignments with the tuned models and selects the least
expensive configuration that satisfies the SLOs. The experiments in this paper
exercise only the replica-count decision on a single accelerator type, so the
accelerator-assignment path is implemented but not evaluated here. Fourth, the actuator translates
the decision into Kubernetes scaling signals.
Two configuration inputs constrain this process. The \emph{staticData} specifies
the available accelerator types and service-class SLOs, while the
\emph{dynamicData} records how many accelerators are currently available.

\section{Comparison with a Decode-Throughput Analyzer}\label{sec:baseline}

We compare our queueing-model analyzer with the decode-throughput analyzer from
the llm-d workload variant autoscaler (WVA)~\cite{llm-d-wva}. We implement its
replica-sizing logic in our control loop. Both arms use the same collector and
actuator, while each analyzer reads the metrics required by its own calculation.
The comparison therefore isolates the two analyzers and does not evaluate the
complete WVA system.

The analyzer estimates how many decode tokens each replica can produce per second
at a target KV-cache utilization. It then compares the replicas' combined capacity
with the decode-token demand from arriving and queued requests and adjusts the
replica count when the two differ.

To estimate supply at the target utilization, the analyzer first models
inter-token latency as a linear function of KV-cache utilization $\kvu$
\begin{equation}\eqnlabel{ta-itl}
    \widehat{\itl}(\kvu) = a\,\kvu + b ,
\end{equation}
where $\kvu \in [0,1]$. Each observation is taken in one control cycle $t$ and pairs a
measured utilization $\kvu_t$ with the corresponding ITL $y_t$. The analyzer estimates both
$a$ and $b$ by least squares when its observation window contains at least $N_{\min}$ points
and satisfies
$\max_t \kvu_t-\min_t \kvu_t \geq \Delta_{\kvu}$. This utilization-range condition
prevents the analyzer from estimating a slope from points clustered near the
same utilization. We call this two-parameter estimate the unconstrained fit.

The window retains at most $N_{\max}$ observations and discards points older
than $\tage$ or outside $[\kvlo,\kvhi]$. The lower utilization bound
excludes near-idle measurements that provide little information about the
slope, while the upper bound excludes measurements taken after the server is
saturated. Because the ITL relationship also depends on request shape, the
analyzer clears the window when the mean input or output length changes by more
than the relative tolerance $\epsshape$.

When the window cannot support the unconstrained fit, the analyzer uses the
replicas' observations from the current control cycle to estimate only the
slope while holding the intercept at $b_{\mathrm p}$. We call this the constrained fit.
Before the first unconstrained fit, $b_{\mathrm p}$ is the hardware baseline
$b_0$, which represents near-zero-load ITL. After an unconstrained fit, the
analyzer retains its fitted intercept as $b_{\mathrm p}$. If neither fit can be
evaluated, our implementation keeps the current replica count.
Configuration parameters of the throughput analyzer and their default values are provided in~\tabref{ta-constants}.

Using this ITL model, the analyzer calculates decode-token supply. For mean input
and output lengths $\il$ and $\ol$, and a hit rate $h$ for the reuse of cached prefixes
shared across requests,
the analyzer estimates the average KV-cache footprint of each active request as
\begin{equation}\eqnlabel{ta-kvreq}
    \kvreq = \il(1-h) + \frac{\ol}{2}.
\end{equation}
The factor of one half accounts for the KV cache growing during decoding. At the
target utilization $\ksat$, a replica with space for $\kvmax$ tokens supports
\begin{equation}\eqnlabel{ta-nsat}
    N_{\text{sat}} = \frac{\ksat\,\kvmax}{\kvreq},
\end{equation}
active requests and supplies decode tokens at the rate
\begin{equation}\eqnlabel{ta-supply}
    C_{\text{rep}} = \frac{N_{\text{sat}}}{\widehat{\itl}(\ksat)} .
\end{equation}
When $\widehat{\itl}$ is expressed in seconds per token,
$C_{\text{rep}}$ is the decode capacity of one replica in tokens per second.
Let $r$ denote the number of ready replicas and $q$ the provisioned replica count.
The corresponding capacities are $C_{\text{ready}}=rC_{\text{rep}}$ and
$C_{\text{prov}}=qC_{\text{rep}}$, also in decode tokens per second. The analyzer
uses provisioned capacity to avoid repeated scale-up requests while replicas start
and ready capacity to prevent replicas that are not yet serving from creating a
scale-down signal.

The analyzer also checks whether the decode rate implied by the fitted model agrees
with the rate reported by the serving engine. If their relative difference exceeds
$\epsrate$ for $\nver$ consecutive control cycles at utilization
$\kvu\geq\kvver$, it clears the observation window and uses the
constrained fit in the next cycle. The last fitted intercept is retained.

The demand estimate has an arrival term and a queued-work term. At arrival rate $\lam$ each request
eventually emits $\ol$ tokens. A standing queue of $Q_w$ requests is assigned a
drain horizon of $\phi\,\widehat{\itl}(\ksat)\ol$ seconds. Multiplying the queued
requests by $\ol$ tokens and dividing by this horizon cancels $\ol$, giving
\begin{equation}\eqnlabel{ta-demand}
    D = \lam\,\ol + \frac{Q_w}{\phi\,\widehat{\itl}(\ksat)} .
\end{equation}
Here $\lam$ is measured in requests per second, so $D$ is decode-token demand in
tokens per second.

The thresholds $\thup$ and $\thdn$ are dimensionless fractions of
capacity. The products $\thup C_{\text{prov}}$ and
$\thdn C_{\text{ready}}$ are therefore demand thresholds, both measured
in tokens per second. The analyzer first tests whether
$D>\thup C_{\text{prov}}$ and scales up if provisioned capacity is
insufficient. Only when this test fails does it consider scale-down. It removes
replicas if $D<\thdn C_{\text{ready}}$ and otherwise keeps the current
allocation. Thus, the scale-down branch below requires both that no scale-up is
pending and that ready capacity has sufficient slack. Our single-variant
implementation selects
\begin{equation}\eqnlabel{ta-decision}
    q' = \begin{cases}
        \max\!\left(1,\left\lceil\dfrac{D}{\thup C_{\text{rep}}}\right\rceil\right),
        & D > \thup C_{\text{prov}}, \\[6pt]
        \max\!\left(1,q-r+\left\lceil\dfrac{D}{\thdn C_{\text{rep}}}\right\rceil\right),
        & \substack{D \leq \thup C_{\text{prov}},\\ D < \thdn C_{\text{ready}}}, \\[6pt]
        q, & \text{otherwise.}
    \end{cases}
\end{equation}
The $q-r$ term preserves provisioned replicas that are not yet ready. The lower
bound of one replica belongs to our actuation harness,
which cannot reactivate a deployment from zero without the ITL and KV-utilization
measurements produced by a serving replica.

The decode-throughput analyzer uses neither a TTFT nor
an ITL SLO target. The fitted ITL calibrates decode-token supply rather than serving
as a latency constraint. Input length affects $\kvreq$ and therefore the estimated
supply, but the analyzer does not translate input length into prefill latency. It
accounts for an existing backlog through \eqnref{ta-demand}, but does not predict
queueing delay or TTFT.

The decode-throughput analyzer depends on fifteen configuration parameters across calibration, consistency checks,
demand estimation, and scaling. WVA supplies these values as defaults, so users
need not configure them individually. Nevertheless, the settings interact in
determining when the ITL model is fitted or reset and when replicas are added or
removed. We retain the pinned upstream defaults without tuning them for either
scenario. \tabref{ta-constants} groups the settings by their role.

\begin{table}[t]
  \centering
  \caption{Configuration used for the throughput analyzer, grouped by role. All
  values except $h$ are pinned upstream defaults~\cite{llm-d-wva}. We set $h=0$
  by disabling prefix caching in both arms.}
  \label{tab:ta-constants}
  \begin{tabularx}{\columnwidth}{@{}lX@{}}
    \hline
    Role & Parameters and values \\
    \hline
    ITL fit & $b_0=6$~ms, $N_{\min}=10$, $N_{\max}=20$,
      $\Delta_{\kvu}=0.30$, $\tage=30$~min,
      $[\kvlo,\kvhi]=[0.15,0.85]$ \\
    Fit reset & $\epsshape=0.20$, $\epsrate=15\%$, $\nver=3$,
      $\kvver=0.30$ \\
    Supply and demand & $\ksat=0.85$, $\phi=2.0$, $h=0$ \\
    Scaling & $\thup=0.85$, $\thdn=0.70$ \\
    \hline
  \end{tabularx}
\end{table}

Our implementation reproduces the decode-throughput analyzer described above with the
configuration in \tabref{ta-constants}. During each control cycle, the evaluator
reads ITL from vLLM's histogram counters, while the collector obtains KV-cache
utilization, waiting-request count, cache capacity, and the cumulative number of
generated tokens from vLLM. The offered arrival rate is the rate configured at the
load generator.
The analyzer records one deployment-level observation per cycle. It averages
utilization across replicas, sums their waiting-request counts and generated-token
increments, and computes a throughput-weighted mean ITL. The ready count $r$
counts ready replicas whose metric scrapes succeed, while $q$ is the last replica count
written by the actuator. The cache capacity $\kvmax$ is the value reported by vLLM
when it initializes the cache.
Because the throughput analyzer performs its own calibration in
\eqnref{ta-itl}, the parameter tuner of \secref{estimation} is inactive in the
baseline arm.
The analyzer returns only a replica count. We retain the current accelerator
assignment and batch limit so that the comparison changes only the replica-sizing
decision.

Both arms serve Llama-3.1-8B with vLLM 0.21.0 on an OpenShift cluster of H100
GPUs, using one GPU per replica. Prefix caching is disabled, and vLLM reports
$\kvmax\approx449{,}200$ KV-cache tokens per replica. Each run uses a 30-second
measurement interval and the 44-minute load profile in
\tabref{baseline-phases}. A control cycle spans this measurement window together with
metric collection and actuation, so one cycle occupies about 42 seconds of wall-clock time
rather than the 30 seconds of the window alone. Within a scenario, both arms use the same controller
image, configuration, and load-generator seed. Runs stop when the wall-clock
profile ends rather than after a fixed cycle count, producing 63 or 64 complete
control cycles per arm. We set the maximum batch size to $B=256$ in vLLM and both
analyzers, matching the offline validation in \secref{validation}. Because vLLM
admits at most $B$ active requests and \eqnref{ta-kvreq} assigns each request an
average footprint of $\kvreq$ tokens, the KV-cache utilization that the analyzer
itself models cannot exceed $\min\{1,B\kvreq/\kvmax\}$, which is $0.73$ for the
TTFT-bound request shape and $0.66$ for the ITL-bound one. These are ceilings on
the analyzer's own estimate rather than on the measured occupancy, since
\eqnref{ta-kvreq} charges every request the time-averaged footprint $\ol/2$ while
requests late in generation hold more. The measured utilizations reported below
can therefore sit slightly above the corresponding modeled ceiling.

\tabref{baseline-scenarios} summarizes the two scenarios. The long-prompt
scenario is TTFT-bound. The short-prompt, long-generation scenario is ITL-bound
and therefore evaluates the throughput analyzer on the metric calibrated by
\eqnref{ta-itl}.

\begin{table}[t]
  \centering
  \caption{The two comparison scenarios. SLO targets were chosen by evaluating the model of \secref{analysis} so that the intended metric binds within the load range.}
  \label{tab:baseline-scenarios}
  \begin{tabular}{lrr}
    \hline
                                  & TTFT-bound & ITL-bound \\
    \hline
    input length $\il$            & $1024$   & $128$   \\
    output length $\ol$           & $512$    & $2048$  \\
    $\ttft$ SLO (ms)              & $50$     & $200$   \\
    $\itl$ SLO (ms)               & $25$     & $13$    \\
    arrival-rate range (RPM)      & $300$--$1200$ & $120$--$480$ \\
    max batch size $B$            & $256$    & $256$   \\
    \hline
  \end{tabular}
\end{table}

\begin{table}[t]
  \centering
  \caption{The 44-minute load profile. Rates are the scenario-specific baseline
  rates of \tabref{baseline-scenarios} multiplied by the indicated factor.}
  \label{tab:baseline-phases}
  \begin{tabular}{lrr}
    \hline
    Segment & Duration & Rate multiplier \\
    \hline
    Baseline hold & 5 min & $1.0$ \\
    Ramp up & 3 min & $1.0\to2.5$ \\
    Mid-load hold & 8 min & $2.5$ \\
    Ramp up & 3 min & $2.5\to4.0$ \\
    Peak hold & 8 min & $4.0$ \\
    Ramp down & 3 min & $4.0\to2.5$ \\
    Mid-load hold & 6 min & $2.5$ \\
    Ramp down & 3 min & $2.5\to1.0$ \\
    Baseline hold & 5 min & $1.0$ \\
    \hline
  \end{tabular}
\end{table}

\begin{figure*}[t]
  \centering
  \includegraphics[width=\textwidth]{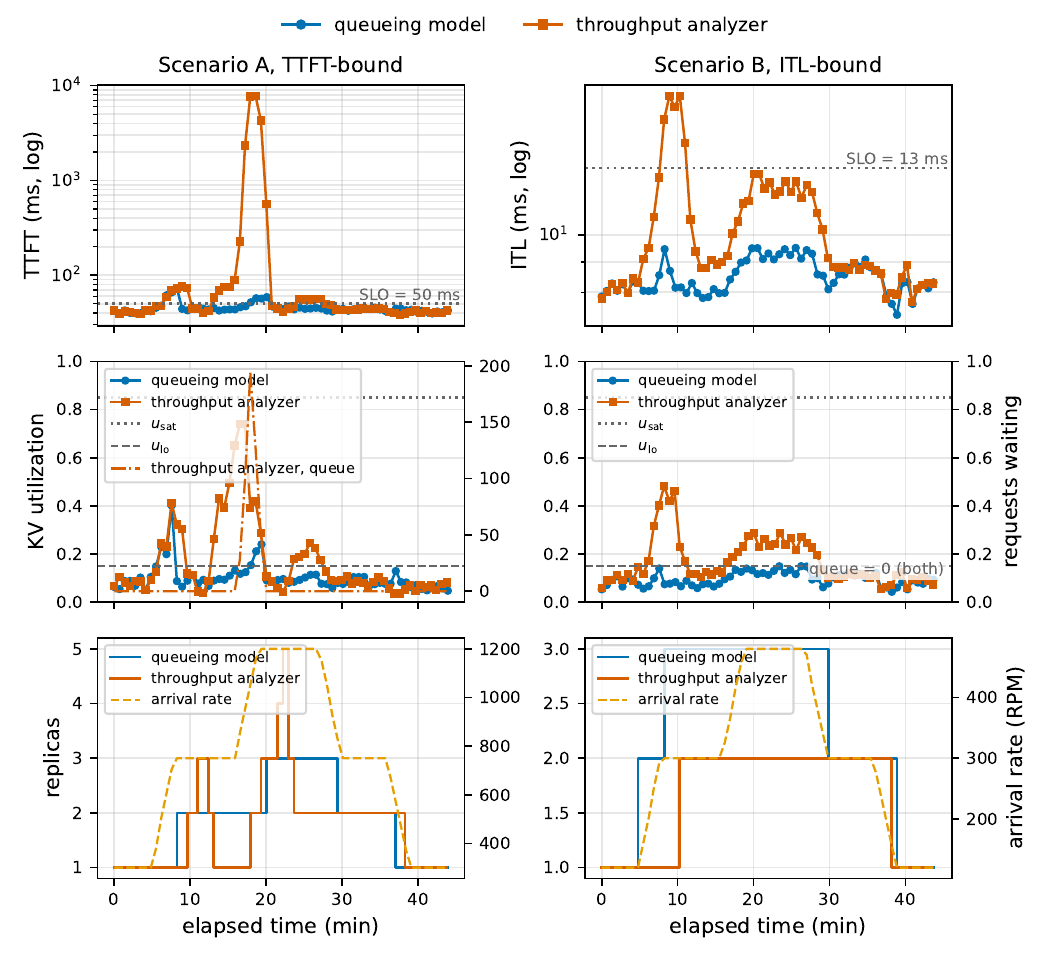}
  \caption{Analyzer comparison on a real H100 vLLM deployment. Left column is the TTFT-bound scenario and right column the ITL-bound one. Top row is the binding latency metric against its SLO, middle row is KV-cache utilization with queue depth on the right axis, and bottom row is the ready replica count with offered arrival rate on the right axis. In the TTFT-bound scenario, the throughput arm's queue buildup coincides with its TTFT excursion. Both queues remain empty in the ITL-bound scenario.}
  \label{fig:analyzer-comparison}
\end{figure*}

During these runs, the throughput analyzer used its unconstrained fit in 27 of 64
TTFT-bound cycles and used the constrained fit throughout the ITL-bound scenario.
The fitting band is the reason it relied on the constrained fit so often. Only 21 of
the 64 TTFT-bound cycles and 26 of the 64 ITL-bound cycles produced a utilization
sample inside $[\kvlo,\kvhi]$, since most cycles ran below $\kvlo$, and the ITL-bound
samples span only $0.32$ in total, barely more than the $\Delta_{\kvu}$ spread that the
unconstrained fit requires.
\Figref{analyzer-comparison} shows both scenarios, and
\tabref{baseline-results} summarizes their outcomes.

\begin{table}[t]
  \centering
  \caption{Analyzer comparison. Latencies are cycle means in milliseconds. Medians and 90th percentiles are across control cycles. Replica counts are provisioned replicas.}
  \label{tab:baseline-results}
  \begin{tabular}{lrr}
    \hline
    & queueing model & throughput analyzer \\
    \hline
    \multicolumn{3}{l}{\emph{TTFT-bound scenario, TTFT SLO $=50$~ms}} \\
    mean provisioned replicas   & $1.88$      & $1.81$        \\
    peak provisioned replicas   & $3$         & $5$           \\
    TTFT median                 & $43.8$      & $44.5$        \\
    TTFT 90th percentile        & $50.6$      & $85.1$        \\
    TTFT worst                  & $73.4$      & $7{,}767$     \\
    cycles over TTFT SLO        & $7/64$      & $22/64$       \\
    cycles over ITL SLO ($25$)  & $0/64$      & $3/64$        \\
    KV utilization range        & $0.05$--$0.40$ & $0.04$--$0.74$ \\
    peak queue depth            & $0$         & $194$         \\
    \hline
    \multicolumn{3}{l}{\emph{ITL-bound scenario, ITL SLO $=13$~ms}} \\
    mean provisioned replicas   & $2.27$      & $1.64$        \\
    peak provisioned replicas   & $3$         & $2$           \\
    ITL median                  & $8.3$       & $9.0$         \\
    ITL 90th percentile         & $9.3$       & $12.7$        \\
    ITL worst                   & $9.5$       & $17.2$        \\
    cycles over ITL SLO         & $0/63$      & $5/64$        \\
    cycles over TTFT SLO ($200$)& $0/63$      & $0/64$        \\
    KV utilization range        & $0.04$--$0.15$ & $0.06$--$0.48$ \\
    peak queue depth            & $0$         & $0$           \\
    \hline
  \end{tabular}
\end{table}

The results show a capacity--latency tradeoff. In the TTFT-bound scenario, the
queueing-model analyzer exceeds the target in $7$ of $64$ cycles, compared with
$22$ of $64$ for the throughput analyzer. In the ITL-bound scenario, the
corresponding counts are $0$ of $63$ and $5$ of $64$. Across both scenarios the
queueing-model analyzer therefore misses a target in $7$ of its $127$ cycles against $27$
of $128$, the union in each arm coinciding with its TTFT-bound count. The throughput analyzer
uses $4\%$ and $28\%$ fewer mean replicas, $1.81$ versus $1.88$ in the
TTFT-bound scenario and $1.64$ versus $2.27$ in the ITL-bound scenario. These comparisons use cycle-mean latency, not
tail latency within a cycle.

The middle row of \Figref{analyzer-comparison} shows the operating regimes behind
these outcomes.
In the TTFT-bound scenario, the queueing-model arm reaches KV utilization $0.40$
with no queued requests, whereas the throughput arm reaches $0.74$ and a queue of
$194$ requests. That queue buildup coincides with the large TTFT excursion. In the
ITL-bound scenario, both queues remain empty, but the throughput arm reaches KV
utilization $0.48$ while the queueing arm remains at or below $0.15$. Thus queueing
explains the largest excursion in the TTFT-bound scenario but not the five
exceedances in the ITL-bound scenario. In the latter case, the throughput analyzer runs with fewer replicas and higher
occupancy, which increases batch iteration time even without a standing queue.
The ITL-bound scenario also shows the difference between calibrating with ITL and
constraining ITL. The throughput analyzer uses the fitted ITL to calculate
decode-token capacity in \eqnref{ta-supply}, whereas the queueing-model analyzer
uses the ITL target as a constraint when choosing replicas.

The runs also provide a closed-loop accuracy check. Against measurements from the
two queueing arms, the median relative errors of predicted TTFT are $3.0\%$ and
$4.7\%$, and those of predicted ITL are $4.3\%$ and $8.2\%$, for the TTFT-bound
and ITL-bound scenarios, respectively. These runs use the same $B=256$ as the offline validation, while
the online tuner refreshes the parameters from the observations gathered under
each fixed request shape.

Each arm is a single run, so the replica and violation counts carry no variance estimate,
and repeated trials are left to future work. Two things bound what that costs. The
closed-loop accuracy figures above are per-cycle comparisons across the $127$ control
cycles of the two queueing arms rather than one measurement each. The difference in
outcomes is also traced to an identified mechanism, namely the queue that forms once the
throughput arm drives KV utilization to $0.74$, rather than inferred from the size of the
gap alone. Moreover, we
evaluate the upstream decode-throughput analyzer inside our collector and direct
actuator, not the complete WVA system. Consequently, our experiment obtains the
arrival rate from the load generator and the queue length from vLLM rather than
from WVA's routing and scheduling components. The comparison therefore
characterizes these two analyzers under the stated harness and defaults, not every
deployment or configuration of the upstream system.

\section{Discussion and Future Work}\label{sec:discussion}

The current model deliberately represents every request by the workload's mean
input and output lengths. This keeps the Markov chain one-dimensional because its
state records only the number of requests in the system. Our validation varies
individual lengths uniformly within $[x/2,3x/2]$ of each configured mean, but the
closed-loop comparison uses one fixed input--output length pair per scenario. It
therefore does not establish accuracy for workloads with wide or multimodal length
distributions, where short and long requests may share the same batch. A
trace-driven evaluation can test this boundary directly. If the mean-length
approximation becomes inaccurate, the state can be extended with a small number of
sequence-length classes rather than one class per request length. The same
extension could use the occupancy probabilities $\{\pi_i\}$ to predict latency
distributions and allow the controller to enforce percentile rather than only mean
TTFT and ITL targets.

The validation also covers light to moderate load under Poisson arrivals. We omit
the runs nearest saturation because their queueing estimates require longer traces
to stabilize. In that regime, the predicted delay is also more sensitive to the
mean-occupancy approximation in \secref{analysis:markov}, which replaces the
occupancy distribution with its mean $\Avg{X}$. Longer experiments
near saturation, followed by bursty and production-trace arrivals, would show when
this approximation ceases to be adequate. They would also reveal whether arrival
variability must become an explicit model input instead of representing demand by
its mean rate alone.

Our analytical model replaces the occupancy distribution and request-length
mixture with mean values. A discrete-event simulator would test the effect of
these approximations by using the same measured per-iteration costs while
explicitly replaying each request arrival, batch admission, prefill chunk, decode
step, and departure. It would provide a higher-fidelity model of queue and batch
evolution without requiring a new GPU experiment for every candidate allocation.
A controlled study could compare both approaches with held-out GPU traces and with
a black-box regressor trained on the same measurements. Prediction error,
robustness after workload shifts, and evaluation time per candidate allocation
would show when the analytical model's compact form is preferable to the
additional detail or flexibility of these alternatives.

The analyzer comparison also identifies a specific control-design issue. In the
ITL-bound scenario, neither analyzer observes a waiting queue, yet the
decode-throughput analyzer operates at higher KV-cache utilization and exceeds the
ITL target in five cycles. Larger active batches increase iteration time before a
queue forms. Consequently, using measured ITL to calculate decode capacity does
not ensure that a future allocation will satisfy an ITL target. A hybrid analyzer
could retain the throughput estimate while rejecting allocations whose predicted
TTFT or ITL exceeds the corresponding target.

We held the decode-throughput analyzer's fifteen settings at their upstream
defaults and continuously re-estimated the queueing model's three parameters. The
current experiments do not show how errors or alternate choices in either set
affect scaling decisions. A sensitivity study should vary the calibration-window
requirements, reset criteria, utilization targets, and queue-drain factor alongside
perturbations of $\al$, $\bt$, and $\gam$. This would identify which settings
materially affect each analyzer and whether a smaller configuration interface can
retain the same behavior.

Finally, the closed-loop comparison uses one model--accelerator pair and one run
per scenario. Repeating each profile would quantify run-to-run variation, while
additional models, accelerators, and request-length distributions would test how
well the conclusions generalize. Both analyzers also respond to the measured
arrival rate. A forecast-driven controller could instead evaluate the capacity
needed one replica-startup delay ahead and begin loading weights before an expected
increase in traffic.

\section{Conclusion}\label{sec:conclusion}

We developed a tractable, parametrized queueing model of an LLM inference server.
It predicts the two primary latency metrics, $\ttft$ and $\itl$, from three
model--accelerator-specific parameters, namely $\al$, $\bt$, and $\gam$.
The model combines mean-value analysis of per-iteration work with a state-dependent
Markov chain over the active batch size and supports both unlimited and chunked prefill.
We validated the model against measurements from a vLLM~\cite{vllm,Kwon2023} server running Llama-3.1-8B and Qwen2.5-14B on an H100 GPU, obtaining ITL relative errors below $8 \%$ and TTFT relative errors below $16 \%$ over a grid of input and output token distributions and request rates spanning light to moderate load.
We estimated the parameters from observed latencies using both an initial,
multi-observation Nelder-Mead fit and an online sliding-window fit that refreshes at
every control cycle.
We then built an SLO-driven autoscaling controller that combines the model tuner
with a queueing-model analyzer and deployed it on an H100 vLLM cluster.
We compared it against a decode-throughput analyzer under one ramping load profile
applied at two scenario-specific rate ranges, while holding the rest of the control loop
fixed. The
throughput analyzer provisioned fewer replicas on average in both scenarios. The
queueing-model analyzer recorded $7/64$ TTFT-SLO violations versus $22/64$ for the
throughput analyzer in the TTFT-bound scenario, and $0/63$ ITL-SLO violations versus
$5/64$ in the ITL-bound
scenario. These results show a capacity--latency tradeoff under the evaluated
workloads rather than universal dominance of either analyzer. The tradeoff is asymmetric in
one respect. The decode-throughput analyzer accepts no latency target, so its replica
counts follow from sizing decode capacity rather than from meeting a stated TTFT or ITL
bound, whereas the queueing-model analyzer treats both targets as constraints.
At the time of writing, our queue analyzer and model tuner are available in experimental mode as components of the llm-d project~\cite{llm-d}, where they support SLO-aware inference orchestration.

The extensions outlined in \secref{discussion} would broaden the workload and
operating regimes covered by the model while preserving its role as a tractable
predictor inside the control loop.

\bibliographystyle{abbrv}
\bibliography{cite}

\appendices
\section{Mean-field approximation of prefill iterations}\label{apx:chunks}

We use a conditional mean-field approximation to determine the effective number
of equal-sized prefill iterations at a fixed total occupancy $i$. This quantity,
denoted $\nc(i)$, is an analytical service-time approximation rather than the
literal number of chunks selected by the runtime scheduler. Let $\mbt$ be the
maximum number of tokens processed per iteration, and condition on one tagged
request being in one of its prefill iterations. The other $i-1$ active requests
form the background population.

Under the homogeneous-request approximation, a background request spends $\nc$
effective iterations in prefill and $\ol$ iterations in decode. At an iteration
boundary, its mean token contribution is therefore
\begin{equation*}
    \frac{\nc}{\nc+\ol}\frac{\il}{\nc}
    + \frac{\ol}{\nc+\ol}
    = \frac{\il+\ol}{\nc+\ol}.
\end{equation*}
We replace the random aggregate contribution of the background requests by its
expectation and approximate each prefill iteration of the tagged request by
$\il/\nc$ tokens. The resulting conditional mean token load is
\begin{equation}\eqnlabel{budget}
    \tload_i(\nc)
    = (i-1)\frac{\il+\ol}{\nc+\ol} + \frac{\il}{\nc}.
\end{equation}
The mean-field token-budget constraint is $\tload_i(\nc)\leq\mbt$. Because
$\tload_i(\nc)$ is strictly decreasing for $\nc>0$, the smallest feasible integer is
obtained from the unique positive solution of $\tload_i(\nc)=\mbt$. Multiplying by
$\nc(\nc+\ol)$ yields
\begin{equation*}
    \mbt\nc^2 + \Phi_i\nc - \il\ol = 0,
    \qquad
    \Phi_i = \ol(\mbt-i+1)-i\il.
\end{equation*}
Hence the positive real-valued threshold is
\begin{equation}\eqnlabel{nc}
    \ncmin(i)
    = \frac{-\Phi_i + \sqrt{\Phi_i^2+4\mbt\il\ol}}{2\mbt},
\end{equation}
and the effective prefill-iteration count used at occupancy $i$ is
\begin{equation}\eqnlabel{nc-solution}
    \nc(i) = \left\lceil \ncmin(i) \right\rceil.
\end{equation}
The analysis assumes $1\leq i\leq B\leq\mbt$, so at least one token per active
request can be accommodated. Since
$\sqrt{\Phi_i^2+4\mbt\il\ol}>|\Phi_i|$, the threshold in \eqnref{nc} is
strictly positive and \eqnref{nc-solution} is at least one.

The limiting cases provide useful checks. With only the tagged request present
($i=1$), equations \eqnref{nc} and \eqnref{nc-solution} reduce to
$\nc(1)=\lceil\il/\mbt\rceil$. For long outputs, the background population is
decode-dominated and consumes approximately $i-1$ tokens per iteration, giving
$\nc(i)\approx\lceil\il/(\mbt-i+1)\rceil$. For long inputs,
$\Phi_i\approx-i\il$ and
$\nc(i)\approx\lceil i\il/\mbt\rceil$, reflecting competition among all $i$
active requests for the token budget.

\section{The unlimited case as a special case}\label{apx:unlimited}

From equations \eqnref{nc} and \eqnref{nc-solution}, when $\mbt \to \infty$,
$\ncmin(i) \to 0^+$ and therefore $\nc(i) = 1$. The state space then reduces to
$\Acal = \{0, 1, \ldots, \ol\}$ with $\ol+1$ equally likely states.
\Eqnref{wpref} reduces to $\wpref\big|_{\nc=1} = (\bt+\gam)\il$, equation \eqnref{wtotal} to $\wtot\big|_{\nc=1} = \bt(\il+\ol) + \gam(\ol+1)(\il+\ol/2)$, and equation \eqnref{delta} to $\delta\big|_{\nc=1} = \bt(\il+\ol)/(\ol+1) + \gam(\il+\ol/2)$.
Within the model, setting $\nc=1$ saves $\gam(\nc-1)\il/2$ of repeated KV reads,
so the unlimited case minimizes latency. Chunking is imposed only when $\mbt$
cannot fit an $\il$-token prefill.
This ordering follows from the model averaging each request's work over its
iterations. A larger $\nc$ lowers $\delta$ and therefore each individual iteration
time, but it raises the total work $\wtot$ and hence the service time $\tau(i)$. What
the averaging does not represent is the single long iteration that a full-prompt
prefill produces inside a mixed batch, which is the practical motivation for
chunking, so the model should not be read as an argument against it.

\end{document}